\documentclass[%
preprint,
 amsmath,amssymb,
 aps,superscriptaddress
]{revtex4-1}

\usepackage{color}
\usepackage{graphicx}
\usepackage{bm}

\graphicspath{{/home/dino/Documents/Specific/Paper/Figures/}}
\begin{document}


\title{Effect of Grafting on Aggregation of Intrinsically Disordered Proteins}

\author{Dino Osmanovic}
\email{d.osmanovic@ucl.ac.uk}
\affiliation{Department of Physics, and Institute of Nanotechnology and Advanced Materials, Bar-Ilan University, Ramat-Gan 52900, Israel}
\author{Yitzhak Rabin}%
\affiliation{Department of Physics, and Institute of Nanotechnology and Advanced Materials, Bar-Ilan University, Ramat-Gan 52900, Israel}
\affiliation{NYU-ECNU Institute of Physics at NYU Shanghai, 3663 Zhongshan Road North, Shanghai, 200062, China}
\date{\today}

\begin{abstract}
A significant part of the proteome is composed of intrinsically-disordered proteins (IDPs). These proteins do not fold into a well-defined structure and behave  like ordinary polymers. In this work we consider IDPs which have the tendency to aggregate, model them as heteropolymers that contain a small number of associating monomers and use computer simulations in order to compare the aggregation of such IDPs that are grafted to a surface or free in solution. We then discuss how such grafting may affect the analysis of in-vitro experiments and could also be used to suppress harmful aggregation.
\end{abstract}

\pacs{Valid PACS appear here}
\maketitle

\section{Introduction}
Proteins are usually thought of as having a well-defined tertiary structure (unique native state) that is determined by their amino-acid sequence. In the simplest models of protein folding\cite{pande_2000-UJ}, proteins fold into this spatial structure based on the sequence of hydrophobic and hydrophilic amino-acids. Because hydrophobic monomers attract and hydrophylic monomers repel each other, the former are found in the interior and the latter are found on the surface of folded proteins\cite{dobson_2004-Fd}. This guarantees the solubility and suppresses the aggregation of proteins in the aqueous environment of the cell. The above strategy fails in the case of IDPs which do not fold into a well-defined three dimensional structure, but still have a functional role to play in the cell\cite{dunker_2008-6W,vanderlee_2014-4N} (IDPs that consist of folded domains connected by hydrophylic linkers are an exception to this rule since their hydrophobic parts are not exposed\cite{sandhu_2009-FB}). One such example is the nucleoporins (nups) of the nuclear pore complex (NPC)\cite{beck_2017-cm,denning_2003-x_,allen_2001-CJ}. The NPC controls the transport of macromolecules between the nucleus and the cytoplasm. The nups are block copolymers comprised a folded domain that forms the rigid scaffold of the NPC and an unfolded domain which is enriched in hydrophobic phenylalanine-glycine repeats (FG nups). A coarse-grained representation of the NPC corresponds to FG nups grafted to the inner surface of a channel\cite {Mario_2013}.  FG nups  have alternating hydrophobic and hydrophilic domains along their backbone, which raises the probability of association amongst them when they are in an aqueous environment. While there is agreement on the importance of associations between FG nups and transport receptors\cite {Schulten_2005, Ma_2016}, it is unclear how big a role attractive interactions between nucleoporins plays in the NPC\cite{osmanovi_2013-AZ}, with one school of thought asserting that the nucleoporins form a strong hydrogel and another stating that they are mainly brush-like\cite {Orit_2010}. Another example of such a system are neurofilaments which are important for maintaining axon structure in neurons\cite{yuan_2012--t,deek_2013-4f}. They also contain flexible side arms which respond dynamically to the environment\cite{jayanthi_2013-3U} and associate via electrostatic interactions between oppositely charged residues\cite{beck_2012-oT}. While attractive interactions between the disordered subdomains play an important role in stabilizing the structure of the neurofilament network, aggregation of neurofilaments is also associated with neurodegenerative diseases\cite{alchalabi_2003-Qi}. 

Note that  in both the NPC and the neurofilaments, the IDPs are grafted to a surface. Whilst grafting can provide clear benefits for localization and introduce a degree of spatial ordering in the system, a less explored aspect is how aggregation of IDPs is affected by the grafting. In cases where the IDPs need to associate in order to perform their function, grafting may affect the partitioning of bound monomers between intra and intermolecular bonds and  suppress the formation of harmful larger aggregates. Furthermore, the fact that the chains (in the NPC and in neurofilaments) are grafted \textit{in vivo}, but experiments that test the behavior of IDPs are often done on free chains in solution \textit{in vitro}, could potentially mean that, while the \textit{in vitro} experiments can give us some level of understanding, the propensity to aggregate may be different in the two systems and lead one to erroneous conclusions about \textit{in vivo} behavior. For instance, experiments on nucleoporins in solution suggest that they form hydrogels \cite{labokha_2013-7f,milles_2013-OM}. How this is affected by grafting has not been explored so far.
Taking such ideas as our inspiration, in this paper we will examine the differences between model systems of disordered polymers when they are either free or grafted. In particular we shall investigate how the aggregation of soluble heteropolymers with a small number of attractive groups is affected by the grafting constraint. 

\section{Model system}
The model system we choose is a system of $M$ polymer chains each of $N=50$ monomers (beads). Most of the beads on this chain interact via the repulsive part of a Lennard-Jones potential, and the backbone of the chain is connected via the FENE potential (see SI).
We designate some of the beads along the backbone of this polymer as being ``stickers'' (in the terminology of ref. \cite{Cates_1986}) that have an attractive Lennard-Jones interaction of strength $\epsilon$  with other such stickers. We then study how the system behaves when the chains are either grafted to a surface or are free to move throughout the volume (the latter case has been previously studied using both mean-field\cite{Semenov_1998} and simulation methods\cite{Khalatur_1996}). In this work we use  Langevin dynamics to simulate the above model of associating polymers.  Simulation details are available in the supplementary information.
\begin{figure}
\begin{center}
\includegraphics[width=160mm]{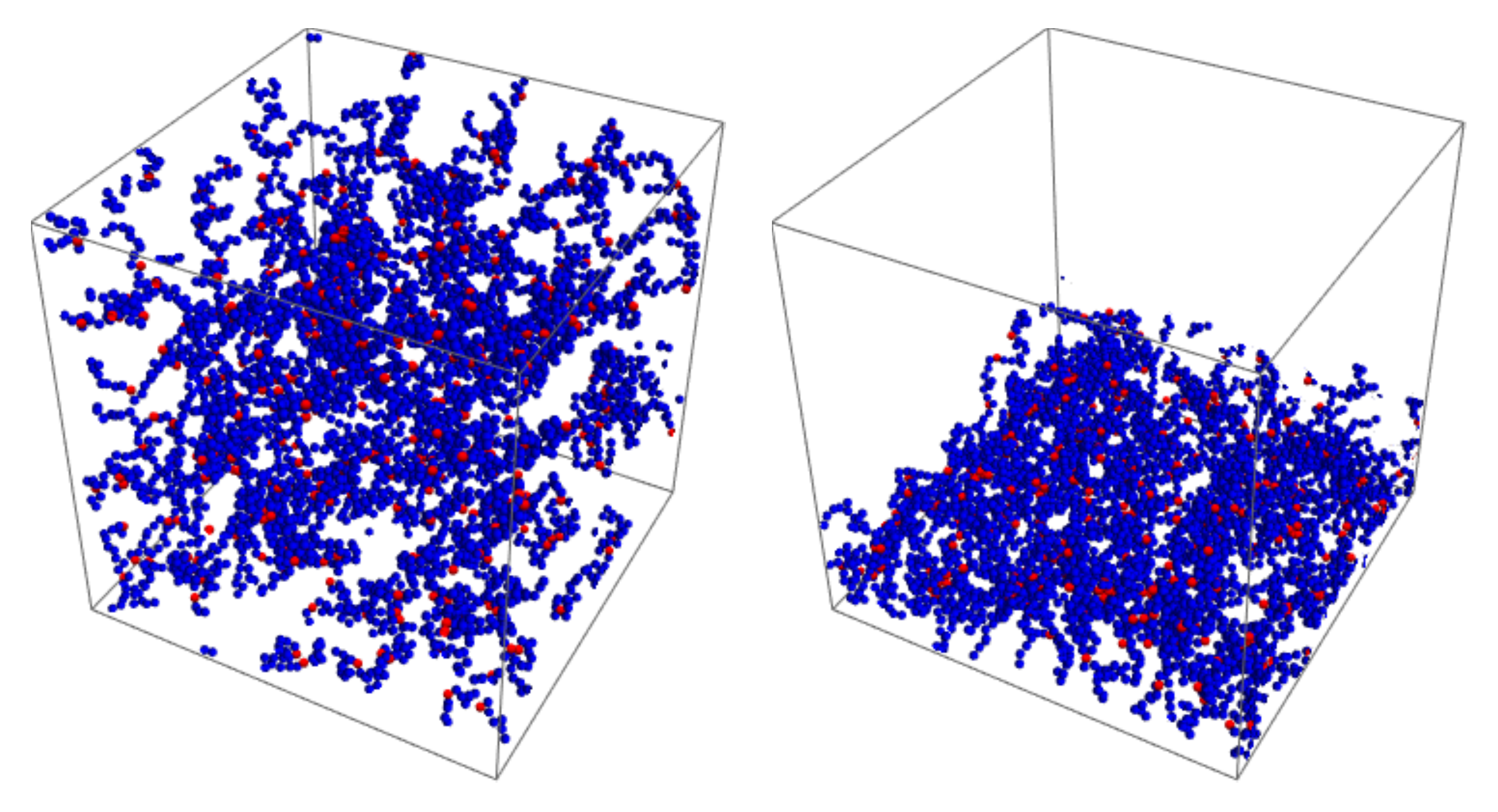}
\caption{Snapshots of the free and the grafted system. The blue beads represent the normal, hard sphere only interaction beads. The red beads are the stickers. The free chains can translate along the system whereas the grafted chains are grafted in a square lattice pattern to the bottom of the system.}
\label{fig:schematic} %
\end{center}
\end{figure}

In fig. \ref{fig:schematic} we show a snapshot of the system for grafted and free chains. There are various parameters we can control in this system: the density $\rho$ of free chains, the grafting density $\rho_g$ of grafted chains, the strength of  interaction between stickers $\epsilon$, the number of stickers on a chain and their positions along the chain contour. We fix the interaction parameter at $\epsilon=5$, and restrict our focus to the cases where the number of stickers is either 3 or 4 (less than 10\% of the beads). This choice guarantees that (a) individual polymers remain soluble and assume expanded conformations in dilute solution and (b) that even though associations are quite strong, they remain reversible in the sense that "bonds" form and break in the course of the simulation. The main quantity we are interested in  studying is how the system behaves as a function of the density. There is a problem here as the two systems have different control parameters which define their density. In the system of free chains, we control the overall density by setting the volume of the simulation. In the system of grafted chains we control the grafting density but the monomer density can vary since it depends on the state of extension of the chains which, in turn, depends on the repulsion between chains, the persistence length, the strength and number of stickers, etc. In order to compare the free and the grafted chains we define the following control parameter $L$ which characterizes the separation between the chains:
\begin{align}
&L=(1/\rho)^{1/3} &\text{mean separation between free chains}\\
&L=(1/\rho_g)^{1/2} &\text{distance between grafting points}
\end{align}

\section{Results}

\begin{figure}[h!]
\begin{center}
\includegraphics[width=180mm]{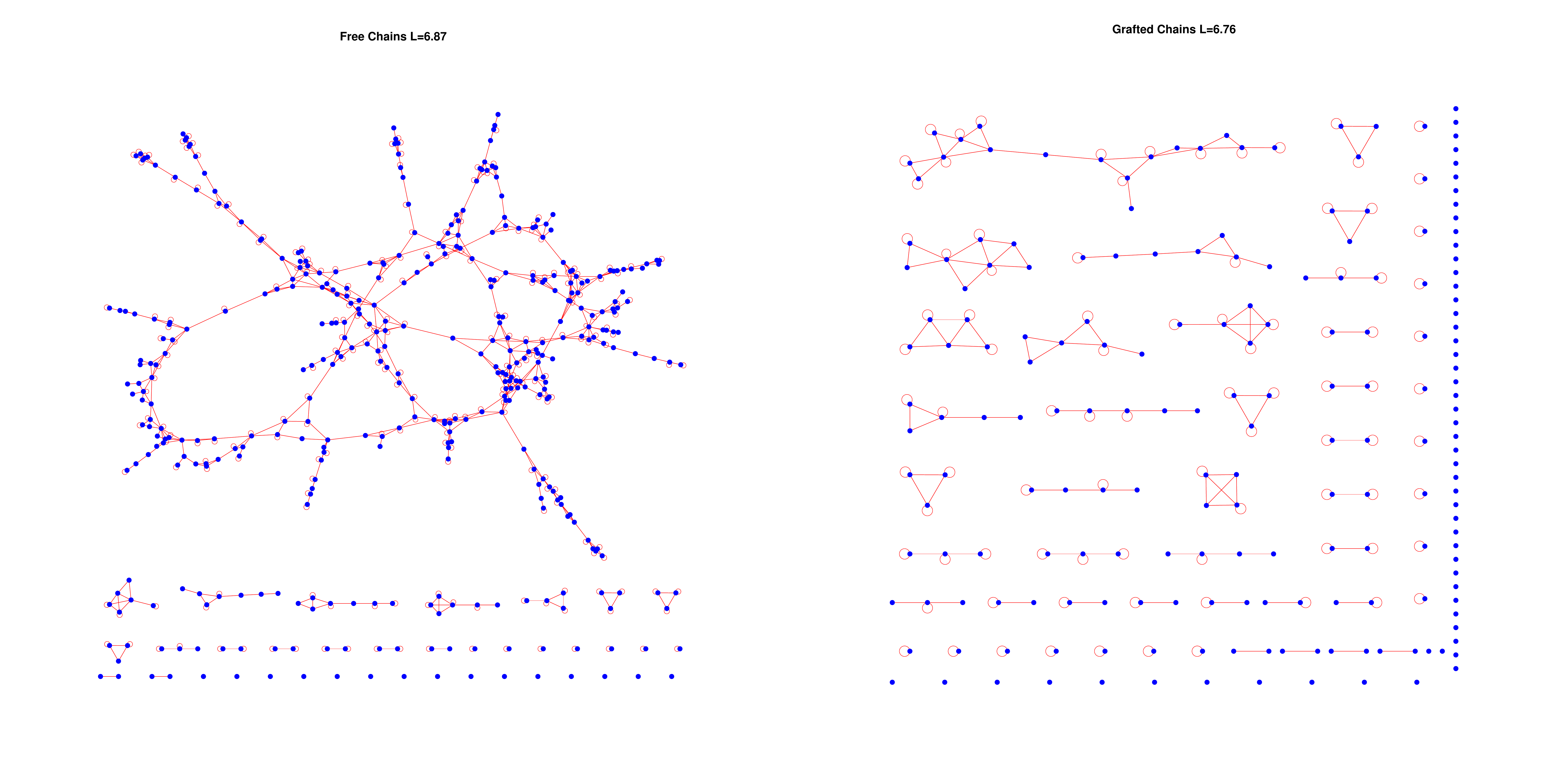}
\caption{An example of graphs of the free and grafted systems. Every blue node represents one of the  polymers. A red edge between the polymers means that the two polymers are bound by at least one pair of stickers. A loop means that the polymer is bound to itself. Both these cases are for 4 stickers per chain and $L\approx7$}
\label{fig:graph} %
\end{center}
\end{figure}

We first look at how the association of the polymers changes as a function of the parameter $L$ for the free and the grafted chains. In order to do this we need a suitable definition of what it means when we say that two polymers are bound together. We choose to define two polymers as being bound together when any of their attractive groups are within a cutoff distance of $1.5$ (in units of the bead diameter $\sigma$ - see SI) from each other. As there are multiple attractive groups per polymer, larger structures can form from these pairwise interactions, \textit{i.e.} when polymer A is bound to polymer B which is bound to polymer C, there is a polymer cluster of size 3. We can categorize all of these pairwise interactions on a graph of the polymers, where an edge between any two polymers indicates that the two are bound. An example of this can be seen in fig. \ref{fig:graph} for both free and grafted chains. Chains can be either bound to other chains, to themselves (represented as loops) or not bound to anything at all. Every snapshot of the system will have some configuration of this sort.

Using these graphs we can analyze the properties of the systems of free and grafted chains. A pertinent question when discussing gelation is the percolation of bound polymer chains in the system. In other words, what proportion of the total number of chains are found in the largest cluster of the system $M_{LC}/M$ (where $M_{LC}$ is the number of polymers in the largest cluster)? If this number is $1$ that means that all the chains in the system are bound in the same cluster (a gel), whereas if it is $0$ that means none of the chains are bound. We can study this ratio for both the grafted and the free systems as a function of $L$. We look at this for both the case where there are 3 and 4 attractive monomers per chain.
\begin{figure}[h!]
\begin{center}
\includegraphics[width=90mm]{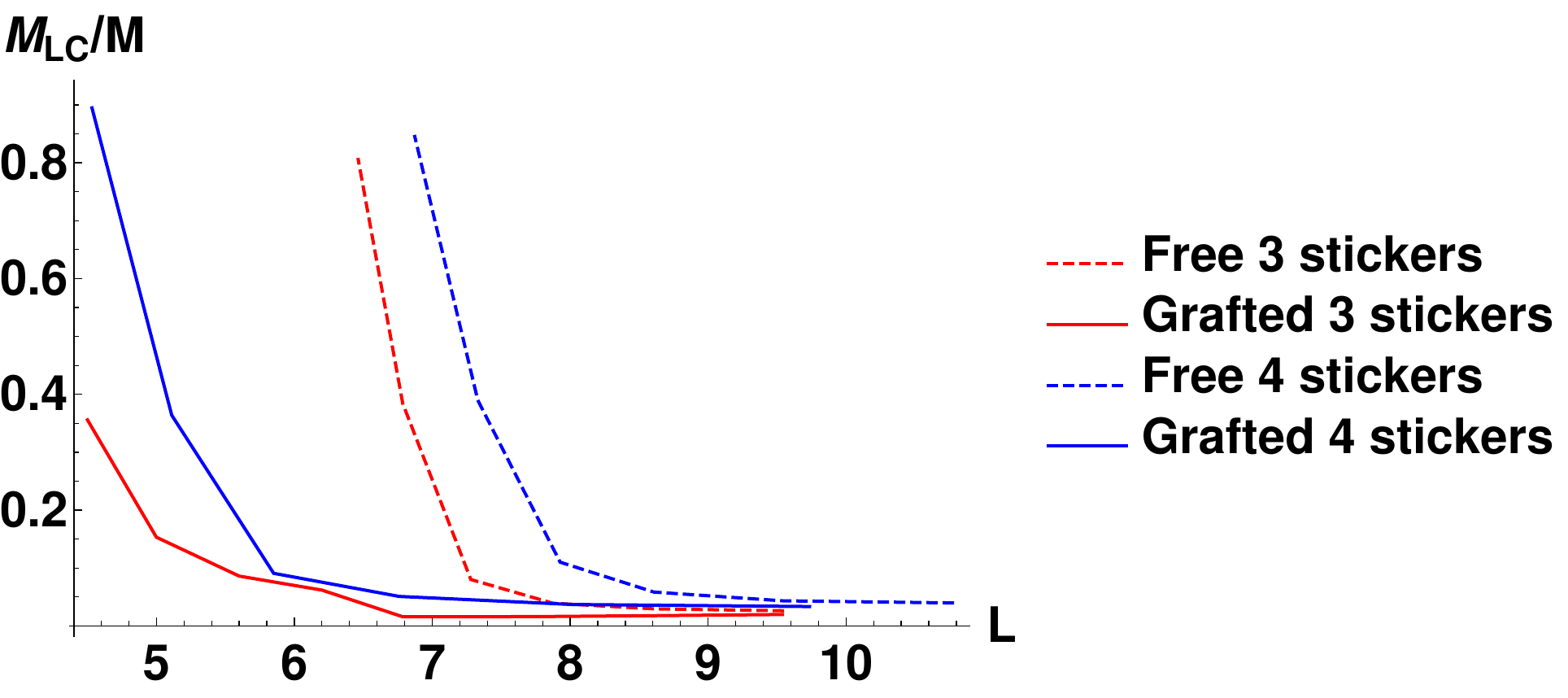}
\caption{The size of the largest cluster as a proportion of the total number of polymers in each system for chains with 3 and 4 attractive points.}
\label{fig:clus3} %
\end{center}
\end{figure}



As seen in fig. \ref{fig:clus3}, there is a large difference in when gelation occurs for grafted and free systems as a function of $L$. As expected, when there are 4 attractive beads the polymers associate into one big cluster at larger values of $L$ than when there are 3 beads. However, the most interesting aspect is when the transitions occur for the grafted and the free chains. When the chains are free the value of $L$ required for the transition to one large interconnected cluster is significantly higher than for the grafted chains. Note that under conditions when the free chains form a connected cluster that percolates through the system, only small localized clusters are observed for grafted chains (fig. \ref{fig:graph}). This concurs with the observation of ``bundles'' in simulations of grafted FG nups\cite{Miao_2009}.

The metric, $M_{LC}/M$, does not provide complete information about the complexity of the system since the two largest clusters in the free and the grafted system may be of the same size but could have significantly different distributions of edges or vertex arrangements. For example  consider the question whether the bonds in the system are intra-chain (two attractive beads on the same chain bound together) or inter-chain (two attractive beads bound that belong to two different polymers) 
\begin{figure}[h!]
\begin{center}
\includegraphics[width=90mm]{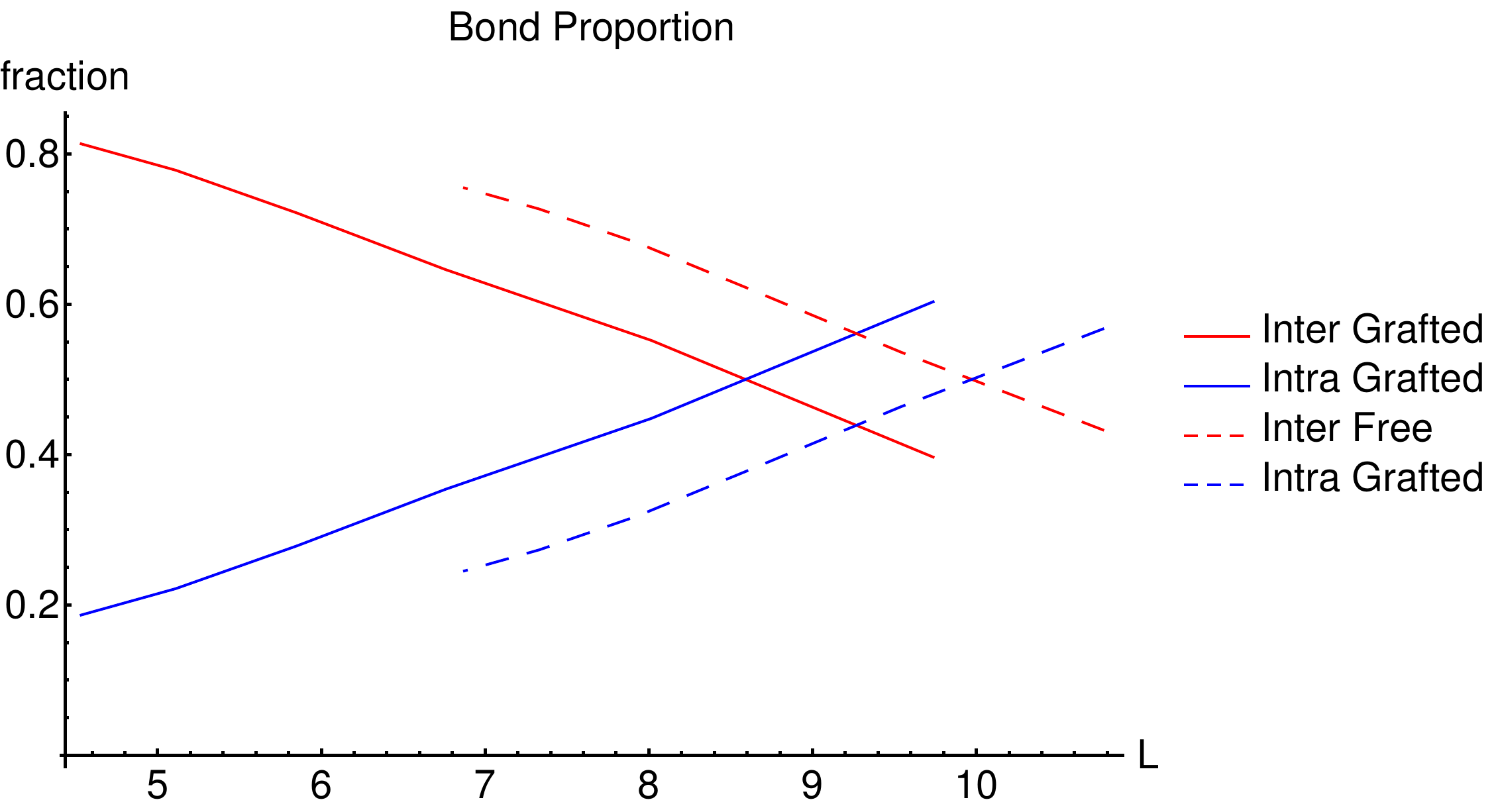}
\caption{The fraction of the total number of bonds that are either inter or intra bonds for polymers with 4 stickers, for both the grafted and free cases. Both have a transition from mainly intra to mainly inter bonds, but this transition occurs at larger $L$ for the free chains. }
\label{fig:degree} %
\end{center}
\end{figure}
In fig. \ref{fig:degree} we observe that both the free and the grafted polymers follow a similar pattern for the proportion of bonds that are intra-chain or inter-chain. As expected, the fraction of intra-chain bonds decreases and that of inter-chain bonds increases with increasing concentration. The main difference is a shift in the value of $L$ where the cross-over from intra to inter-chain association occurs.
\begin{figure}[h!]
\begin{center}
\includegraphics[width=180mm]{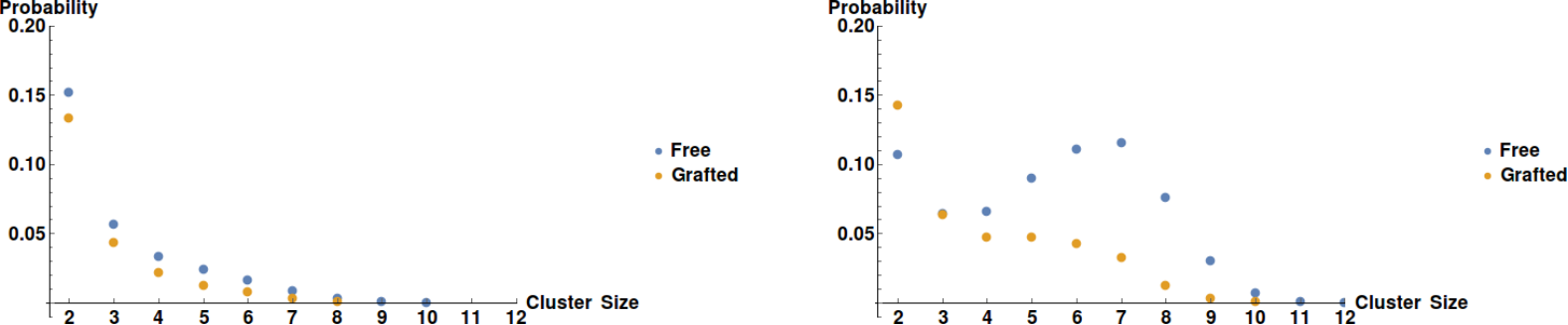}
\caption{The probability that a sticker is found in a cluster that contains a given number of stickers. Distributions are shown for both free and grafted systems at $L=6.75$ and $L=9.5$. Beyond the gel transition of the free system, there is a maximum for at intermediate cluster sizes (for the same value of $L$ a shoulder is observed in the grafted case).}
\label{fig:sticker_cluster} %
\end{center}
\end{figure}
Since in our model the attractive interactions between stickers are non-saturating (there is no imposed limit on the size of clusters of stickers), it is interesting to compare the numbers of stickers in clusters of grafted and free chains. In fig.\ref{fig:sticker_cluster} we show the number of stickers per cluster for the two systems for $L=9.5$ (below gel point concentration for both free and grafted chains ) and for $L=6.75$ (above gel point for free but not for grafted chains). While below percolation threshold both distributions fall monotonically with cluster size as the formation of large clusters of stickers is suppressed by excluded volume repulsions between the chains attached to the stickers in the clusters, a maximum in the distribution of cluster sizes (at about 7 stickers per cluster) is observed above the gel point for the free case. This correlates strongly with the lifetimes of each cluster as seen in figure S1 in the SI which also exhibit a maximum at 8-9  stickers per cluster. The existence of this maximum reflects the interplay of the attraction between stickers (that favors larger clusters) and excluded volume repulsions and loss of conformational entropy (both increasing with cluster size).


\section{Discussion}

In this work we used computer simulations to study the simplest model of free and grafted IDPs - that of a heteropolymer consisting mostly of repulsive and some attractive monomers. In order to model water-soluble IDPs the parameters were chosen such that (a) the radii of gyration of isolated polymers are only weakly perturbed by the presence of associating monomers and (b) the stretching of individual grafted chains is similar to that of a polymer brush in good solvent (not shown). Note that
for the value of the interaction parameter ($\epsilon=5$) and the range of concentrations in our study, if the stickers were not parts of otherwise repulsive polymers, they would be far into the solid regime. The fact that the stickers are short segments of otherwise repulsive polymers surppresses aggregation and limits the maximum possible size of clusters to about 12-14 stickers. This is due to several factors. Firstly, the hard sphere repulsion of the monomers attached to the stickers reduces the second virial coefficient between the polymers\cite {Cates_1986} and makes it more difficult for the stickers to bind together. In addition, the clustering of the stickers comes at some entropic cost (it limits the conformational space of the polymers). 

While it is not immediately clear that grafting would have a significant effect on aggregation, our results suggest that the formation of large aggregates is strongly suppressed by grafting the IDPs to a surface. Note, however, that comparison between two different systems depends on the conditions at which they are studied and ideally, these conditions should be the same. Unfortunately, there is an inherent ambiguity as to whether the free and the grafted systems should be compared at the same distance between monomers or at the same distance between chains. For free polymers both distances can be controlled by adjusting the volume of the system but for grafted polymers only the distance between the chains $L$ can be controlled by adjusting the grafting density (because chains in the brush can stretch, the distance between monomers is not fixed by the grafting constraint).  A large difference between the agreggation propensities of free and grafted chains is found when the comparison is made at fixed $L$ but only minor differences are observed if the comparison is made at similar monomer concentrations. We would like to emphasize that this ambiguity is not a handicap of our model - it would be present in any experiment that attempts to compare systems of free and grafted polymers.

There are intriguing insights from this model that may be of relevance to biological systems. For instance, in the nuclear pore complex there is some confusion over whether the nucleoporins form either a gel or a brush \textit{in vivo}\cite{osmanovi_2013-AZ}. Experimental studies on free nucleoporins can possibly provide some answers about which of these two outcomes is more likely. However, given the differences observed in the results section, such experiments need to take into account that the gel point may be different if the chains were grafted, depending on the various parameters involved in the problem. There are obviously differences between our model and the real system, which has a curved geometry and many different types of polymers, persistence lengths, interactions, etc. However, it may help to understand some of the observed irregularities in this field.

A related question is why are any intrinsically disordered proteins in many living systems attached to surfaces? Obviously, if there is a need to localize chains somewhere, grafting will ensure this. However, there is another possibility, that grafting of chains may be a mechanism by which biological systems can control the extent and the strength of aggregation. In such unfolded proteins, both hydrophobic and hydrophilic domains are exposed to the aqueous environment, leading to the possibility of aggregation. Large aggregates, either from natively unfolded or misfoled proteins can be biologically harmful. By keeping these proteins grafted to a surface, one can  maintain some functionally important degree of association while suppressing the formation of  aggregates.


\begin{acknowledgments}
This work was supported by grants from the Israel Science Foundation and the I-CORE program of
the planning and budgeting committee.
\end{acknowledgments}

\appendix
\section{Simulation details}

We perform the simulations using molecular dynamics with a Langevin thermostat. The temperature of the system is set to $T=1$. The stickers interact via the Lennard-Jones potential:
\begin{equation}
\phi_{LJ}(r)=4\epsilon \left( (\sigma/r)^{12}-(\sigma/r)^{6}\right)
\end{equation}
where $\sigma=1$ and $\epsilon=5$ for the stickers and for all the other particles we only take the repulsive part of the potential. 

The backbone of the polymer is modelled with the FENE potential:
\begin{equation}
\phi_{\text{FENE}}(r)=-0.5 K R_0^2 \ln\left[1-\left(\frac{r}{R_0}\right)^2\right]
\end{equation}
where we take $K=30$ and $R_0=1.5$.

We verified that at these sets of parameters that the interactions in the system were reversible. We first run the simulation for a few thousand Lennard-Jones times for equilibration and then collect data after this point.

\section{Lifetime of Clusters}
\begin{figure}[h!]
\begin{center}
\includegraphics[width=90mm]{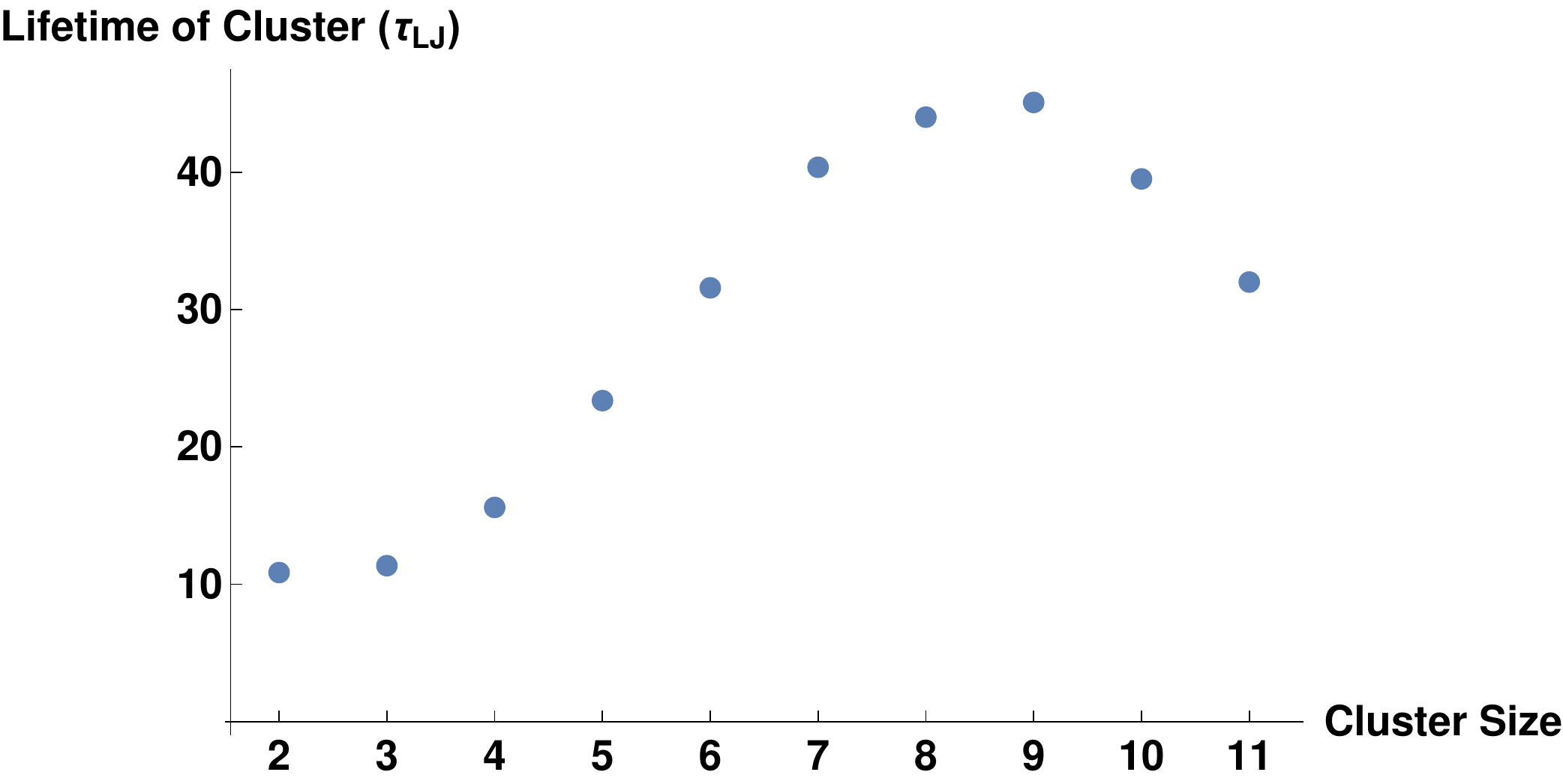}
\caption{The lifetime of a cluster for the free chains when L=6.75, measured by how long one cluster persists before it either loses or gains a particle. This graph looks a lot like the graph of the probability for a particle to be found in a cluster of a given size.}
\label{fig:sticker_cluster} %
\end{center}
\end{figure}
\clearpage

\bibliography{disordered}

\begin{thebibliography}{10}

\bibitem{pande_2000-UJ}
V.~S. Pande, A.~Y. Grosberg, and T.~Tanaka, ``Heteropolymer freezing and
  design: Towards physical models of protein folding,'' {\em Rev. Mod. Phys.},
  vol.~72, pp.~259--314, jan 2000.

\bibitem{dobson_2004-Fd}
C.~M. Dobson, ``Principles of protein folding, misfolding and aggregation.,''
  {\em Semin Cell Dev Biol}, vol.~15, pp.~3--16, feb 2004.

\bibitem{dunker_2008-6W}
A.~K. Dunker, I.~Silman, V.~N. Uversky, and J.~L. Sussman, ``Function and
  structure of inherently disordered proteins.,'' {\em Curr Opin Struct Biol},
  vol.~18, pp.~756--764, dec 2008.

\bibitem{vanderlee_2014-4N}
R.~van~der Lee, M.~Buljan, B.~Lang, R.~J. Weatheritt, G.~W. Daughdrill, A.~K.
  Dunker, M.~Fuxreiter, J.~Gough, J.~Gsponer, D.~T. Jones, P.~M. Kim, R.~W.
  Kriwacki, C.~J. Oldfield, R.~V. Pappu, P.~Tompa, V.~N. Uversky, P.~E. Wright,
  and M.~M. Babu, ``Classification of intrinsically disordered regions and
  proteins.,'' {\em Chem Rev}, vol.~114, pp.~6589--6631, jul 2014.

\bibitem{sandhu_2009-FB}
K.~S. Sandhu, ``Intrinsic disorder explains diverse nuclear roles of chromatin
  remodeling proteins.,'' {\em J Mol Recognit}, vol.~22, pp.~1--8, feb 2009.

\bibitem{beck_2017-cm}
M.~Beck and E.~Hurt, ``The nuclear pore complex: understanding its function
  through structural insight.,'' {\em Nat Rev Mol Cell Biol}, vol.~18, no.~2,
  pp.~73--89, 2017.

\bibitem{denning_2003-x_}
D.~P. Denning, S.~S. Patel, V.~Uversky, A.~L. Fink, and M.~Rexach, ``Disorder
  in the nuclear pore complex: the fg repeat regions of nucleoporins are
  natively unfolded.,'' {\em Proc Natl Acad Sci U S A}, vol.~100,
  pp.~2450--2455, mar 2003.

\bibitem{allen_2001-CJ}
N.~P. Allen, L.~Huang, A.~Burlingame, and M.~Rexach, ``Proteomic analysis of
  nucleoporin interacting proteins.,'' {\em J Biol Chem}, vol.~276,
  pp.~29268--29274, aug 2001.

\bibitem{Mario_2013}
M.~Tagliazucchi, O.~Peleg, M.~Kröger, Y.~Rabin, and I.~Szleifer, ``Effect of
  charge, hydrophobicity and amino-acid sequence of nucleoporins on the
  translocation of model cargoes through the nuclear pore complex,'' {\em
  PNAS}, vol.~110, p.~3363, 2013.

\bibitem{Schulten_2005}
T.~A. Isgro and K.~Schulten, ``Binding dynamics of isolated nucleoporin repeat
  regions to importin-β,'' {\em Structure}, vol.~13, p.~1869–1879, 2005.

\bibitem{Ma_2016}
J.~Ma, A.~Goryaynov, and W.~Yang, ``Super-resolution 3d tomography of
  interactions and competition in the nuclear pore complex,'' {\em Nature
  Structural and Molecular Biology}, vol.~23, pp.~239--247, 2016.

\bibitem{osmanovi_2013-AZ}
D.~Osmanovi{\'{c}}, A.~Fassati, I.~J. Ford, and B.~W. Hoogenboom, ``Physical
  modelling of the nuclear pore complex,'' {\em Soft Matter}, vol.~9, no.~44,
  p.~10442, 2013.

\bibitem{Orit_2010}
O.~Peleg and R.~Y. Lim, ``Converging on the function of intrinsically
  disordered nucleoporins in the nuclear pore complex,'' {\em Biological
  Chemistry}, vol.~391, p.~719–730, 2010.

\bibitem{yuan_2012--t}
A.~Yuan, M.~V. Rao, Veeranna, and R.~A. Nixon, ``Neurofilaments at a glance.,''
  {\em J Cell Sci}, vol.~125, pp.~3257--3263, jul 2012.

\bibitem{deek_2013-4f}
J.~Deek, P.~J. Chung, J.~Kayser, A.~R. Bausch, and C.~R. Safinya,
  ``Neurofilament sidearms modulate parallel and crossed-filament orientations
  inducing nematic to isotropic and re-entrant birefringent hydrogels.,'' {\em
  Nat Commun}, vol.~4, p.~2224, 2013.

\bibitem{jayanthi_2013-3U}
L.~Jayanthi, W.~Stevenson, Y.~Kwak, R.~Chang, and Y.~Gebremichael,
  ``Conformational properties of interacting neurofilaments: Monte carlo
  simulations of cylindrically grafted apposing neurofilament brushes.,'' {\em
  J Biol Phys}, vol.~39, pp.~343--362, jun 2013.

\bibitem{beck_2012-oT}
R.~Beck, J.~Deek, and C.~R. Safinya, ``Structures and interactions in
  'bottlebrush' neurofilaments: the role of charged disordered proteins in
  forming hydrogel networks.,'' {\em Biochem Soc Trans}, vol.~40,
  pp.~1027--1031, oct 2012.

\bibitem{alchalabi_2003-Qi}
A.~Al-Chalabi and C.~C.~J. Miller, ``Neurofilaments and neurological
  disease.,'' {\em Bioessays}, vol.~25, pp.~346--355, apr 2003.

\bibitem{labokha_2013-7f}
A.~A. Labokha, S.~Gradmann, S.~Frey, B.~B. Hülsmann, H.~Urlaub, M.~Baldus, and
  D.~Goerlich, ``Systematic analysis of barrier-forming fg hydrogels from
  xenopus nuclear pore complexes.,'' {\em EMBO J}, vol.~32, pp.~204--218, jan
  2013.

\bibitem{milles_2013-OM}
S.~Milles, K.~Huy~Bui, C.~Koehler, M.~Eltsov, M.~Beck, and E.~A. Lemke,
  ``Facilitated aggregation of fg nucleoporins under molecular crowding
  conditions.,'' {\em EMBO Rep}, vol.~14, pp.~178--183, feb 2013.

\bibitem{Cates_1986}
M.~E. Cates and T.~A. Witten, ``Chain conformation and solubility of
  associating polymers,'' {\em Macromolecules}, vol.~19, no.~3, pp.~732--739,
  1986.

\bibitem{Semenov_1998}
A.~N. Semenov and M.~Rubinstein, ``Thermoreversible gelation in solutions of
  associative polymers. 1. statics,'' {\em Macromolecules}, vol.~31, no.~4,
  pp.~1373--1385, 1998.

\bibitem{Khalatur_1996}
P.~G. Khalatur, A.~R. Khokhlov, I.~A. Nyrkova, and A.~N. Semenov, ``Aggregation
  processes in self-associating polymer systems: Computer simulation study of
  micelles in the superstrong segregation regime,'' {\em Macromolecular Theory
  and Simulations}, vol.~5, no.~4, pp.~713--747, 1996.

\bibitem{Miao_2009}
L.~Miao and K.~Schulten, ``Transport-related structures and processes of the
  nuclear pore complex studied through molecular dynamics.,'' {\em Structure},
  vol.~17, no.~3, pp.~449--459, 2009.

\end{thebibliography}
\bibliographystyle{ieeetr}

\end{document}